\documentclass[10pt, twocolumn, letterpaper]{article}

\usepackage{iccv}
\usepackage{times}
\usepackage{epsfig}
\usepackage{graphicx}
\usepackage{amsmath}
\usepackage{amssymb}

\usepackage{booktabs}
\usepackage{colortbl}
\usepackage{xcolor}
\usepackage{amsmath, bm, subfigure, epstopdf, url, pifont, overpic, cases}
\usepackage{latexsym, amssymb, bbding, multirow, makecell,  diagbox, enumitem, caption}
\usepackage{array, enumitem, soul, algorithm, algpseudocode}

\usepackage{arydshln}

\usepackage{multirow, multicol}
\usepackage{adjustbox}

\newcommand{\name}{0}
\newcommand{\h}{0}
\newcommand{\w}{0.15}
\newcommand{\wa}{0.15}
\newlength \g

\usepackage[pagebackref=true, breaklinks=true, letterpaper=true, colorlinks, bookmarks=false]{hyperref}

\iccvfinalcopy %

\def\iccvPaperID{10} %

\begin{document}

\title{Single Image Super-Resolution Using Lightweight Networks Based on Swin Transformer}
\author{\hspace{-0.4cm}Bolong Zhang$^{1}$ ~ Juan Chen$^{1}$  ~ Quan Wen$^{1}$\\
$^{1}$University Of Electronic Science And Technology Of China\\
}

\maketitle

\begin{abstract}
Image super-resolution reconstruction is an important task in the field of image processing technology, which can restore low resolution image to high quality image with high resolution. In recent years, deep learning has been applied in the field of image super-resolution reconstruction. With the continuous development of deep neural network, the quality of the reconstructed images has been greatly improved, but the model complexity has also been increased. In this paper, we propose two lightweight models named as MSwinSR and UGSwinSR based on Swin Transformer. The most important structure in MSwinSR is called Multi-size Swin Transformer Block (MSTB), which mainly contains four parallel multi-head self-attention (MSA) blocks. UGSwinSR combines U-Net and GAN with Swin Transformer. Both of them can reduce the model complexity, but MSwinSR can reach a higher objective quality, while UGSwinSR can reach a higher perceptual quality. The experimental results demonstrate that MSwinSR increases PSNR by $\mathbf{0.07dB}$ compared with the state-of-the-art model SwinIR, while the number of parameters can reduced by $\mathbf{30.68\%}$, and the calculation cost can reduced by $\mathbf{9.936\%}$. UGSwinSR can effectively reduce the amount of calculation of the network, which can reduced by $\mathbf{90.92\%}$ compared with SwinIR.
\end{abstract}

\begin{figure*}[!htbp]
	\captionsetup{font=small}
	
	\centering
	\scriptsize
	
	\renewcommand{\h}{0.105}
	\renewcommand{\wa}{0.12}
	\newcommand{\wb}{0.16}
	\renewcommand{\g}{-0.7mm}
	\renewcommand{\tabcolsep}{1.8pt}
	\renewcommand{\arraystretch}{1}
	\resizebox{1.00\linewidth}{!} {
		\begin{tabular}{cc}
			
			\renewcommand{\name}{figure/out_imgs_}
			\renewcommand{\h}{0.078}
			\renewcommand{\w}{0.176}
			\begin{tabular}{cc}
				\begin{adjustbox}{valign=t}
					\begin{tabular}{c}
						\includegraphics[height=0.185\textwidth, width=0.185\textwidth]{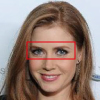}
						\\
						CelebA ($4\times$): 086750.jpg
					\end{tabular}
				\end{adjustbox}
				\begin{adjustbox}{valign=t}
					\begin{tabular}{cccccc}
						\includegraphics[height=\h \textwidth, width=\w \textwidth]{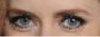} \hspace{\g} &
						\includegraphics[height=\h \textwidth, width=\w \textwidth]{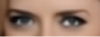} \hspace{\g} &
						\includegraphics[height=\h \textwidth, width=\w \textwidth]{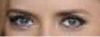} \hspace{\g} &
						\\
						HR \hspace{\g} &
						SwinIR~\cite{liang2021swinir} \hspace{\g} &
						SRGAN~\cite{ledig2017photo} \hspace{\g} 
						\\
						\vspace{-1.5mm}
						\\
						
						\includegraphics[height=\h \textwidth, width=\w \textwidth]{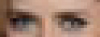} \hspace{\g} &
						\includegraphics[height=\h \textwidth, width=\w \textwidth]{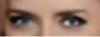} \hspace{\g} &
						\includegraphics[height=\h \textwidth, width=\w \textwidth]{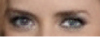}
						\hspace{\g} &	
						\\ 
						
						LR \hspace{\g} &
						MSwinSR (ours)  \hspace{\g} &
						UGSwinSR (ours) \hspace{\g}
						\\
					\end{tabular}
				\end{adjustbox}
			\end{tabular}
			
		\end{tabular}
	}\vspace{-2mm}
	\caption{Comparison of the reconstructed image ($\times 4$) generated by different models. SwinIR\cite{liang2021swinir} and MSwinSR use Swin Transformer with high objective quality. SRGAN\cite{ledig2017photo} and UGSwinSR use GAN with high perceptual quality.}
	\label{fig:out_img_3}
\end{figure*}

\section{Introduction}
Super resolution (SR) is a computer vision and image processing technology that reconstructs a high-resolution (HR) image from its low-resolution (LR) image \cite{wang2020deep, li2021beginner, anwar2020deep}. It can improve the clarity of the displayed image without changing the physical properties of the imaging equipment. Therefore, this technology can be applied to medical imaging\cite{isaac2015super}, security monitoring\cite{zhang2010super}, remote sensing image processing\cite{ji2019vehicle} and other fields, reducing the cost of upgrading imaging equipment due to the need for higher image quality. In addition, it can also be applied to the early stage of data preprocessing\cite{dai2016image}, which can effectively improve the performance of edge detection, semantic segmentation, digit recognition and scene recognition.

Due to the successful performance of convolutional neural network (CNN) in the field of computer vision, it has also been used in the field of SR\cite{dong2014learning, kim2016accurate, kim2016deeply, zhang2018residual, goodfellow2014generative}, besides, the quality of reconstructed images has been greatly improved compared with traditional machine learning algorithms. However, CNN also has some disadvantages, because the convolution kernel can not be well adapted to the content of the image. This deficiency can be improved by replacing CNN with Transformer\cite{vaswani2017attention}, which uses a self-attentional mechanism, capturing global content information of images, showing promising performance in several computer vision tasks\cite{carion2020end, dosovitskiy2020image, liu2021swin, touvron2021training}. However, Transformer can cost large amount of calculation. Recently, Swin Transformer\cite{liu2021swin} has been introduced to reduce the computation, and it can capture local information of image content just like CNN.

SwinIR\cite{liang2021swinir} applied Swin Transformer to SR for the first time. Compared with CNN and networks with attention mechanism, SwinIR has the advantages of fewer parameters and higher objective quality of reconstructed images. But SwinIR also have the following drawbacks: (1) Since capturing the attention mechanism is implemented by the global information of source image, the overall reconstruction image is more smooth, while some local details are difficult to be detected. This has little effect on higher-resolution images, but it will greatly reduce the  perceptual quality of small-size images. (2) Besides the Swin Transformer block, SwinIR also uses a large number of convolutional layers, which will increase the amount of computation in the network. If these convolutional layers are deleted, the reconstruction quality of the image will be greatly reduced. (3) In order to solve the special problem of SR, SwinIR cancels the downsampling operation in Swin Transformer. Although this can reduce the number of parameters, it will also increase the amount of calculation of the model, and it is difficult to extract deeper features of the images.

In this paper, we propose two models to solve the problems mentioned above. The first model named Multi-size Swin SR (MsSwinSR) uses multiple blocks with different attention windows to process feature maps in parallel, so that a single multilayer perceptron (MLP) block can process the information obtained by multiple Swin Transformer blocks simultaneously. Therefore it can reduce the number of MLP blocks, building a lightweight network with both less computation and fewer parameters. In other words, the original network’s depth is reduced, but the width of the network increases. As a result, the performance of MSwinSR doesn’t change greatly compared with SwinIR, since it uses the same count of Swin Transormer blocks which play an important role in SR.

The second model named as U-net GAN Swin SR (UGSwinSR). We add the downsampling operation into SwinIR, so that the deep features of the image could be extracted. However, due to the particularity of SR, upsampling operation is required after downsampling to restore the features. Therefore, we refer to the design of U-net\cite{ronneberger2015u, cao2021swin} and removed the convolutional layer, which greatly reduces the computation of the network. Due to the downsampling operation, the original image information is destroyed, hence this structure can only be used for extracting the deep features. But for the original image, other methods that can restore LR to HR are required. In order to reduce the size of the model, we use BICUBIC\cite{catmull1978recursively}, a simple interpolation, to obtain HR. Generative Adversarial Network (GAN)\cite{goodfellow2014generative} is also used in this model for higher perception quality. The results show that we can obtain promising image perception quality with very low computational cost.

\section{Related Work}
\label{sec:headings}

\subsection{Image Super-Resolution}
Image super-resolution was first proposed by Harris\cite{harris1964diffraction} in the 1960s. Early image super-resolution technologies are mainly based on interpolation methods, such as nearest neighbor interpolation, bilinear interpolation and bicubic interpolation\cite{catmull1978recursively}. With the continuous development of machine learning, Freeman et al.\cite{freeman2000learning} introduced machine learning to the field of SR for the first time in 2000. Subsequently, a variety of reconstruction algorithms based on machine learning have emerged, such as the algorithm based on neighborhood embedding\cite{chang2004super}, the algorithm based on sparse representation\cite{yang2008image} and the algorithm based on local linear regression\cite{timofte2013anchored}.

However, traditional super-resolution reconstruction algorithms and machine learning algorithms mostly make use of the underlying features of the image, so the reconstruction performance is greatly  limited, besides it is difficult to reconstruct the edge, contour, texture and other details of the high-resolution image. Therefore, in order to extract deep features of images, deep learning algorithm is applied to the field of super-resolution.

Dong et al.\cite{dong2014learning} applied convolutional neural network in the field of image super-resolution in 2014 for the first time. This work inspired researchers to apply neural networks to the field of super-resolution and proposed a large number of super-resolution reconstruction algorithms based on deep learning\cite{kim2016accurate, kim2016deeply, zhang2018residual, goodfellow2014generative}.

\subsection{Swin Transformer}

The attention mechanism was first used in the field of natural language processing to solve the problem that the recurrent neural network\cite{zaremba2014recurrent} could not perform parallel computation and had long-term dependence on sequential information. The most famous model of the attention mechanism is called Transformer\cite{vaswani2017attention}. Transformer is mainly used in the field of natural language processing and has achieved great success. Vision Transformer (ViT)\cite{dosovitskiy2020image} introduced Transformer into the field of Computer Vision for the first time and worked out promising result. Swin Transformer\cite{liu2021swin} is improved on the basis of ViT by introducing sliding window. On the one hand, it has the ability to process local information, and on the other hand, it has less computation compared with ViT. SwinIR\cite{liang2021swinir} uses Swin Transformer model, which is currently performing well in the field of computer vision, and has the best performance among all models in the objective evaluation index of reconstruction effect with relatively few parameters.
\subsection{U-net}
U-net\cite{ronneberger2015u} is widely used in medical image segmentation. U-net includes two modules: feature extraction and upsampling, also known as encoder-decoder. Feature images obtained from feature extraction and upsampling at the same scale need concat operation, which is called skip connection in the network. Cao et al.\cite{cao2021swin} have combined U-net with Swin Transformer for medical image segmentation. Wang et al.\cite{wang2022uformer} use Swin Transformer with a U-shaped structure for image restoration, including SR.
\subsection{GAN}
Generative Adversarial Networks (GAN)\cite{goodfellow2014generative} consists of a generator to generate the image, and a discriminator to determine whether the image is generated by the generator. SRGAN\cite{ledig2017photo} is the first model to use generative adversarial network in SR. The characteristic of SRGAN is that the loss function is not a single loss function, but a weighted sum of several losses, including the per-pixel loss, content loss and adversarial loss, so as to ensure the perceptual quality of the reconstructed image. If Mean Square Error (MSE) loss is used alone in pursuit of a higher Peak Signal-to-Noise Ratio (PSNR), the reconstructed image will produce an unsatisfactorily smooth texture effect.
\section{Method}
\subsection{MSwinSR}

\begin{figure*}[!htbp]
\captionsetup{font=small}%
\scriptsize
\begin{center}
\includegraphics[width=0.8\textwidth]{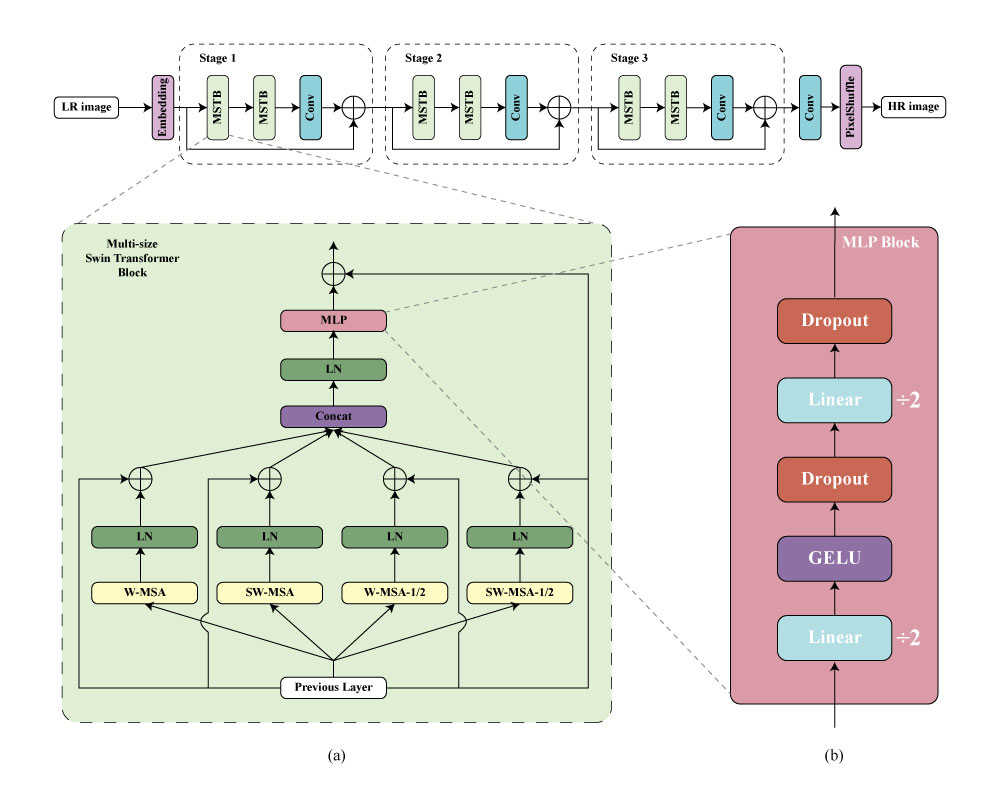}\label{fig:SwinIR_SR}%
\vspace{-0.3cm}
\caption{The architecture of MSwinSR. (a) is the Multi-size Swin Transformer Block. (b) is the MLP Block}\label{fig:MSwinSR}
\end{center}
\vspace{-0.3cm}
\end{figure*}

\begin{figure*}[!htbp]
\captionsetup{font=small}%
\scriptsize
\begin{center}
\includegraphics[width=0.8\textwidth]{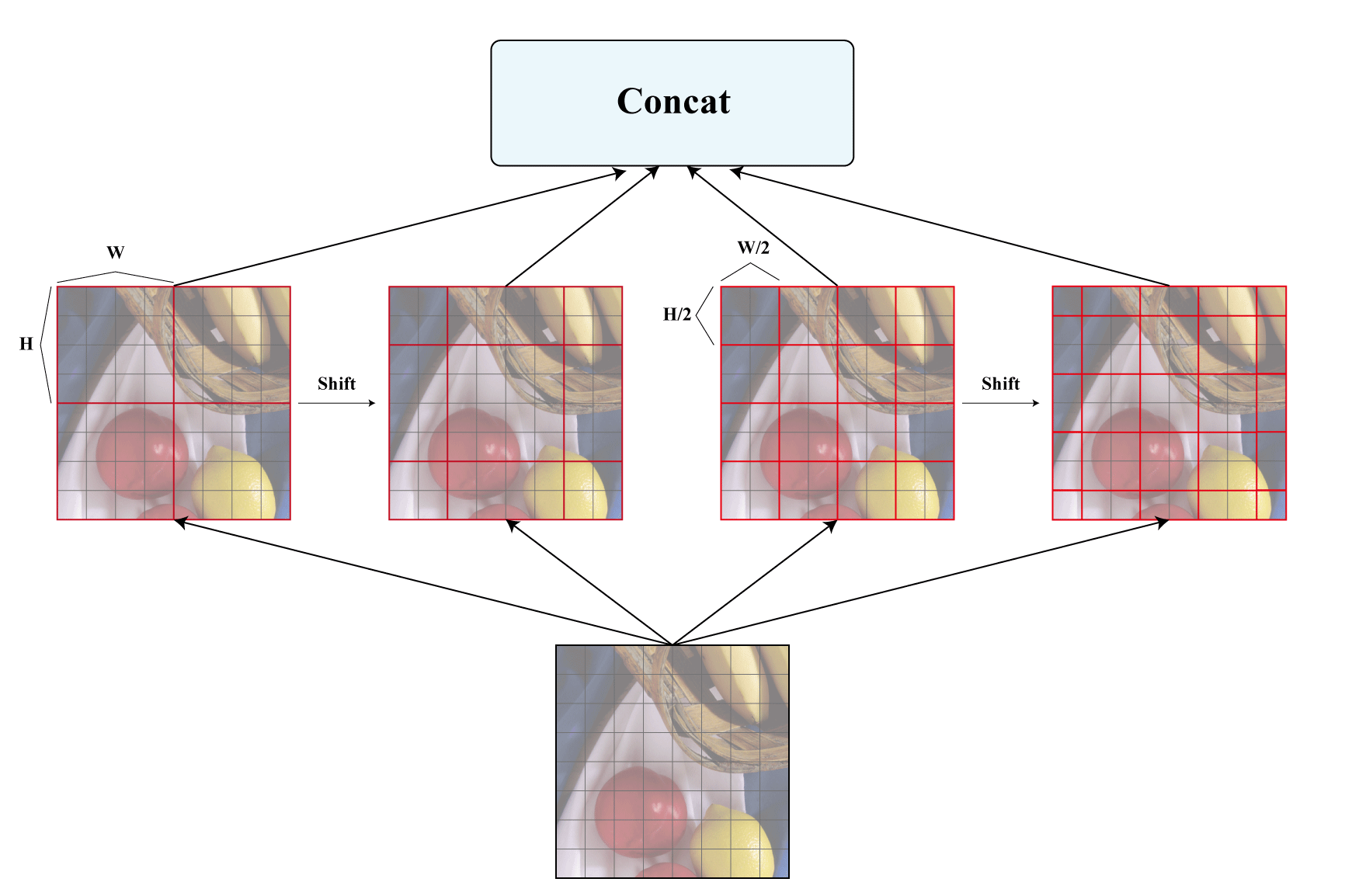}\label{fig:In-Swin}%
\vspace{-0.3cm}
\caption{The different MSA blocks}\label{fig:MSA}
\end{center}
\vspace{-0.3cm}
\end{figure*}

As shown in Figure \ref{fig:MSwinSR}, the network structure of MSwinSR is similar to SwinIR's. The main difference is that the Residual Swin Transformer Block (RSTB) in SwinIR is replaced with Multi-size Swin Transformer Block (MSTB), which mainly contains four parallel multi-head self-attention (MSA) blocks. 

Given a LR image $I_{LR}\in \mathbb{R} ^{H\times W\times C_{in}}$, where $H$, $W$, $C_{in}$ represent the height, weight and channel of the image respectively, we use an Embedding layer to get $F_{0, out}\in \mathbb{R} ^{H\times W\times C}$, the input of MSTB:
\begin{equation}
F_{0, out}=H_{Embed}(I_{LR})
\end{equation}

Where $H_{Embed}(\cdot)$ is the Embedding layer. We use a $3\times 3$ convolutional layer to implement the functions of this Embedding layer. Then we use $K$ Stage, which consists of $L$ MSTB, a convolutional layer and a residual connection, represented by a dashed box in Figure \ref{fig:MSwinSR}, to extract deep feature $F_{iL, out}\in \mathbb{R} ^{H\times W\times C}$ :
\begin{equation}
\begin{aligned}
    &F_{iL, out}=H_{Stage_{i}}(F_{j, out})+F_{(i-1)L, out}, \\&i=1, 2, \dots K, \\&j=0, L, 2L, \dots , KL
\end{aligned}
\end{equation}

\begin{equation}
\label{equ:Stage}
F_{i, out}=
\begin{cases}H_{\mathit{MSTB} _{i}}(F_{i-1, out}), i=1, \dots, L-1\\
H_{Conv}(H_{\mathit{MSTB} _{i}}(F_{i-1, out})), i=L
\end{cases}
\end{equation}

Where $H_{Stage_{i}}(\cdot)$ is the $i$-th Stage, whose operation can be represented by Equation \ref{equ:Stage}.

It is worth noting that we use 'depth' to indicate the number of MSTB used throughout the model. For example, in Figure \ref{fig:MSwinSR}, the depth of the model is denoted as [2, 2, 2], which means after every 2 MSTB there is a convolutional layer and a residual connection, represented as a Stage, and there are 3 same structures in the model.

Finally we use PixelShuffle\cite{pixelshuffle}, represented as $H_{PS}(\cdot)$, to reconstruct the HR image $I_{HR}\in \mathbb{R} ^{(Hs)\times (Hs) \times C_{in}}$, where $s$ represents the scale of image magnification. But before PixelShuffle, we need to use a convolutional layer to change the channel of the feature into $s^2C_{in}$. So the process of getting the HR image is formulated as:
\begin{equation}
I_{HR}=H_{PS}(H_{Conv}(F_{KL, out}))
\end{equation}

For the input feature graph $F_{i, 0}$ of the $i$-th MSTB, the output obtained by feature extraction of the $j$-th MSA is formulated as:
\begin{equation}
F_{i, j}=LN(H_{MSA_{i, j}}\left(F_{i, 0}\right))+F_{i, 0}, \quad j=1, 2, 3, 4
\end{equation}

Where $H_{MSA_{i, j}}(\cdot)$ is the $j$-th MSA in the $i$-th MSTB, then the LayerNorm (LN) layer is added before residual connection.

Four outputs of the MSA would be concatenated before a LN layer and a MLP block. The whole output of the $i$-th MSTB is formulated as:
\begin{equation}
F_{i, out}=MLP(LN(Concat(F_{i, 1}, \cdots, F_{i, 4})))+F_{i, 0}
\end{equation}

As shown in Figure \ref{fig:MSwinSR} (b), the MLP block is different from that used in Swin Transformer, since the channel dimension of the input feature map of MLP block is expanded by 4 times after concatenation. Therefore, the two Linear layers in MLP need to reduce the channel dimension by half, so that the output of the MSTB has the same size as the input.

Four different MSA blocks are named as W-MSA, SW-MSA, W-MSA-1/2, SW-MSA-1/2. Where 'W' and 'SW' represent MSA blocks with regular and shifted windowing configurations respectively, while '1/2' represents the window size of the MSA blocks is half of the regular ones. The attention windows of those MSA blocks are shown in Figure \ref{fig:MSA}.

The pixel loss can directly guide the reconstructed image to approach the real high-resolution image in terms of pixel value. Compared with $L1$ loss, $L2$ loss has greater loss on pixel values with larger differences, and is more likely to accept pixel values with smaller differences. However, this will cause the image to be too smooth. In practice, $L1$ loss shows better performance and convergence than $L2$ loss\cite{2018Fast, 2017Enhanced, Hang2017Loss}. So we use $L1$ loss as the loss function to optimize the parameters
of MSwinSR:

\begin{equation}
    \mathcal{L}_{{L}_{1}}(I_{LR}, I_{HR})=\frac{1}{H W C} \sum_{i, j, k} \left | I_{LR_{i, j, k}}-I_{HR_{i, j, k}} \right | 
\end{equation}

Compared with SwinIR, this design has two benefits. On the one hand, there is one MLP block after each MSA in SwinIR, while the MLP in MSwinSR is after four parallel MSA, so that there are fewer MLP blocks and parameters in MSwinSR. On the other hand, it makes sure that the network can take advantage of multiple features extracted by parallel MSA blocks with different attention windows rather than a single MSA. Even if some of these MSA blocks could not extract useful features, the MLP block can reduce the weight to prevent the network from being inefficient.

\begin{figure*}[!htbp]
\captionsetup{font=small}%
\scriptsize
\begin{center}
\includegraphics[width=0.8\textwidth]{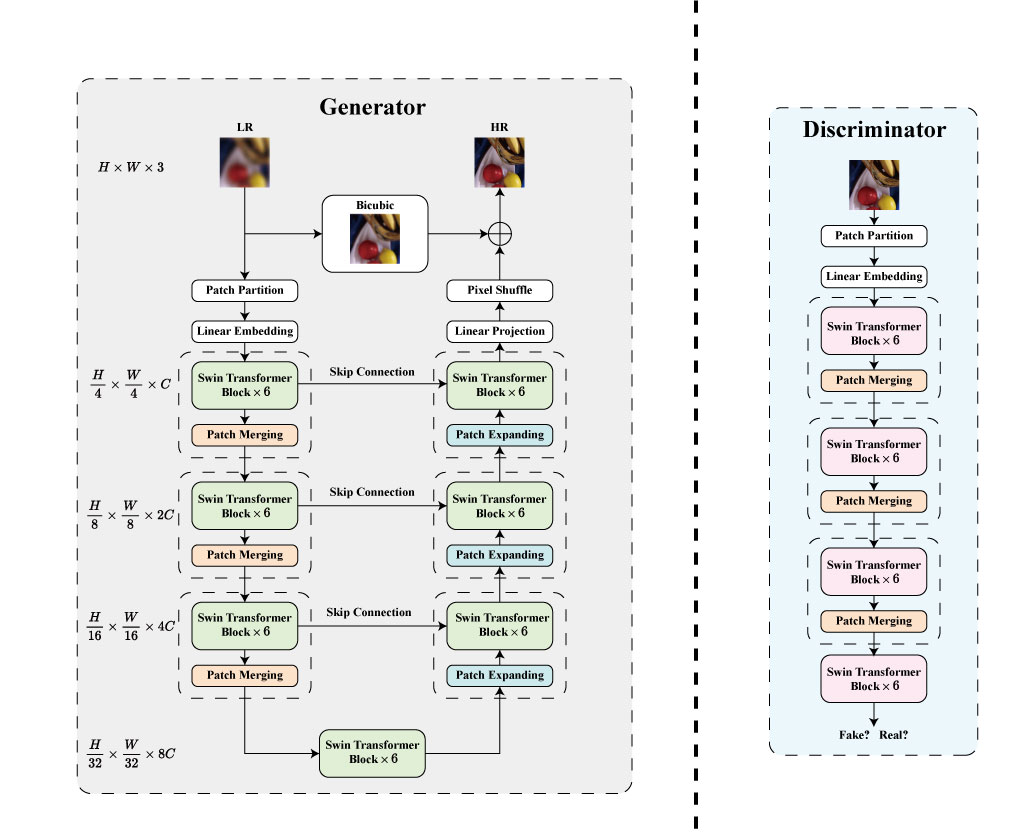}\label{fig:UGSwinSR}%
\vspace{-0.3cm}
\caption{The architecture of UGSwinSR}\label{fig:UGSwinSR}
\end{center}
\vspace{-0.3cm}
\end{figure*}

\subsection{UGSwinSR}
As shown in Figure \ref{fig:UGSwinSR}, UGSwinSR contains two parts: generator and discriminator. The generator can reconstruct HR image from LR image with downsampling and upsampling. While the discriminator only has downsampling since it need to extract the deep features from both the reconstructed images and the ground truth.

Similar to MSwinSR, UGSwinSR also has a parameter 'depth', which represents the times of downsampling. For example, in Figure \ref{fig:UGSwinSR}, the depth of the model is 4.

\subsubsection{Computational Complexity}
The advantage of using U-net is that it can effectively reduce the computational burden of the model. We can compare the RSTB module in SwinIR with the Swin Transformer Block plus Patch Merging module in UGSwinSR which are called a Stage in Swin Transformer\cite{liu2021swin}. Given a LR image as an input, the height and width of the image are $h$ and $w$ respectively. After the Shallow Feature Extraction in SwinIR and the Embedding layer in UGSwinSR, the number of channels in the feature map is $C$. If Softmax and bias are omitted, the computational complexity of W-MSA is\cite{liu2021swin}:

\begin{equation}
\Omega (W\mathrm{-}MSA)=4 h w C^{2}+2 M^{2} h w C
\end{equation}

Six Swin Transformer blocks are included in a RSTB, including a Layer Norm and a MLP in addition to a W-MSA. Therefore, the computational complexity of a RSTB is:

\begin{equation}
\Omega (RSTB)=81 h w C^{2}+12(M^{2}+1) h w C
\end{equation}

While for a Stage in UGSwinSR, taking the first module downsampled by 4 times as an example, a Patch Merging contains a Layer Norm and a fully connected Layer that reduces the number of channels from $4C$ to $2C$, so the computational complexity of a Stage is:

\begin{equation}
\Omega (Stage)= \frac{37}{8}  h w C^{2}+(\frac{3}{4} M^{2}+\frac{13}{16} ) h w C
\end{equation}

\begin{figure*}[!htbp]
\captionsetup{font=small}%
\scriptsize
\begin{center}
\includegraphics[width=0.8\textwidth]{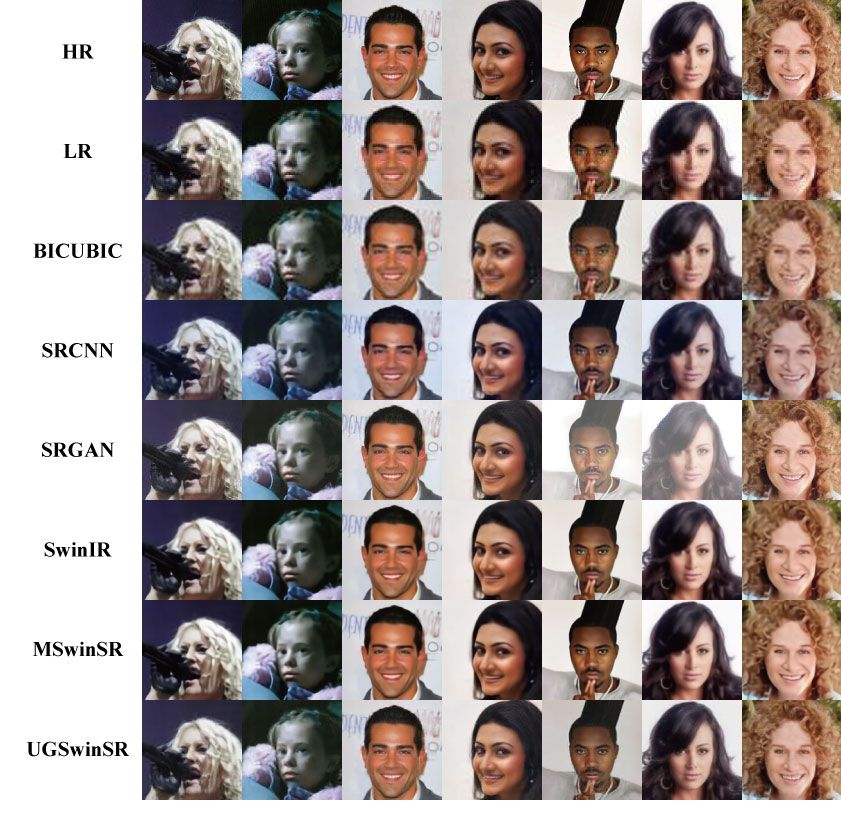}\label{fig:Output_imgs}%
\vspace{-0.3cm}
\caption{Reconstructed image ($\times$4) diagrams generated by different models}\label{fig:Output_imgs}
\end{center}
\vspace{-0.3cm}
\end{figure*}

\section{Experiments}
\subsection{Dataset}
Our models are trained on CelebA\cite{liu2015deep}, a dataset that contains 202599 face images with the size of 178×218. Firstly, an image of 178×178 is cropped in the center of the image, and then the 178×178 image is enlarged to 256×256 image by bicubic interpolation as a HR image. Finally, bicubic downsampling is used as the image degradation, and the 256×256 HR image is scaled to 64×64 LR image, so that the high and low resolution image pairs required for training can be obtained. 10000 images are selected as the training dataset, and another 100 images are selected as the validation dataset.

\begin{figure*}[!htbp]
\captionsetup{font=small}%
\scriptsize
\begin{center}
\includegraphics[width=0.8\textwidth]{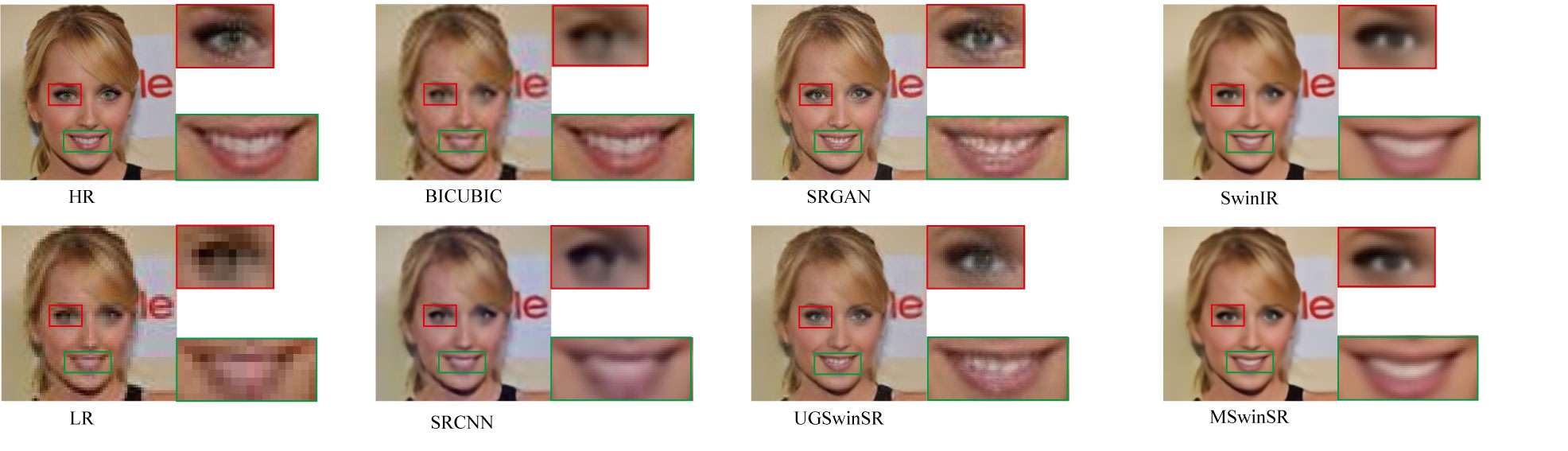}\label{fig:Output_imgs_2}%
\vspace{-0.3cm}
\caption{Detail diagram of reconstructed image generated by different models on a typical image from \cite{liu2015deep}}\label{fig:Output_imgs_2}
\end{center}
\vspace{-0.3cm}
\end{figure*}

\subsection{Experimental Setup}
In the experiment, the training epoch was uniformly set as 100, the batch size was 20 images, the optimizer used for training was Adam\cite{kingma2014adam}, the learning rate was set as 0.0002, and the exponential decay rates were set as 0.5 and 0.999. In the model using the attention mechanism, the number of channels after the embedding layer is set to 60.

\subsection{Results}

Figure \ref{fig:Output_imgs} shows the reconstruction effect of different super-resolution algorithms on the same images, including Bicubic\cite{catmull1978recursively}, SRCNN\cite{2016Image}, SRGAN\cite{ledig2017photo}, SwinIR\cite{liang2021swinir}, MSwinSR and UGSwinSR. 

In Figure \ref{fig:Output_imgs_2}, a typical image in the validation set is taken as an example, and the details of the eyes and mouth are enlarged. In the low-resolution image (lower left), it can be seen that the details of  the eyes and mouth have virtually disappeared, including pupils and teeth. Although all models can basically restore these details, the effects of restoration are different. For example, in SRCNN and SwinIR, the pupil and tooth parts are still of a single color, while the SRGAN and UGSwinSR proposed in this paper can restore the details of reflection inside the pupil and the details of each tooth. In addition, SRGAN is also better for the details of the double eyelid and eye bags, with the highest perception quality.

\begin{table*}[!htbp]
\centering
\caption{Quantitative comparison with different methods. Best and second best performance are in red and blue colors respectively.}
\label{tab:different methods}
\begin{tabular}{lllllll} 
\toprule
Method                                                           & Params~                   & Multi-Adds~               & PSNR~                   & SSIM~                    & LPIPS-Alex~              & LPIPS-VGG~                \\ 
\hline
BICUBIC\cite{catmull1978recursively}                                                          & -                         & -                         & 27.53                   & 0.8222                   & 0.2808                   & 0.2674                    \\
SRCNN\cite{2016Image}~                                                           & \textcolor{red}{69.25k~}  & 4.538G~                   & 27.66                   & 0.8455                   & 0.2048                   & 0.2562                    \\
SRGAN\cite{ledig2017photo}~ ~ ~                                                       & 1.550M                    & 4.791G                    & 24.30                   & 0.7939                   & \textcolor{red}{0.1202}  & \textcolor{red}{0.2100}   \\
SwinIR\cite{liang2021swinir}                                                           & 897.2k~                   & 4.187G~                   & \textcolor{blue}{29.47} & \textcolor{red}{0.8661}  & 0.1834                   & 0.2233                    \\ 
\hdashline
MSwinSR (ours)                                                   & \textcolor{blue}{621.9k}~ & 3.771G~                   & \textcolor{red}{29.54}  & \textcolor{blue}{0.8647} & 0.1892                   & 0.2266                    \\
USwinSR (ours)                                                   & 19.85M~                   & \textcolor{blue}{380.0M~} & 28.08                   & 0.8447                   & 0.1853                   & 0.2293                    \\
UGSwinSR (ours)                                                  & 19.85M~                   & \textcolor{blue}{380.0M~} & 26.48                   & 0.8146                   & \textcolor{blue}{0.1268} & \textcolor{blue}{0.2101}  \\
\begin{tabular}[c]{@{}l@{}}UGSwinSR+MSwinSR\\(ours)\end{tabular} & 11.42M~                   & \textcolor{red}{255.3M~}  & 26.28                   & 0.8164                   & 0.2139                   & 0.2602                    \\
\bottomrule
\end{tabular}
\end{table*}

\begin{table*}[!htbp]
\centering
\caption{Quantitative comparison with different depths in MSwinSR and UGSwinSR. The best performance in each model is in red color.}
\label{tab:depths}
\begin{tabular}{llllllll} 
\toprule
Method                                                                    & depth                & Params~  & Multi-Adds~ & PSNR~                   & SSIM~                    & LPIPS-Alex~              & LPIPS-VGG~               \\ 
\hline
\multirow{3}{*}{\begin{tabular}[c]{@{}l@{}}MSwinSR \\(ours)\end{tabular}}
                                                                          & {[}1, 1, 1, 1, 1]    & 701.1k   & 3.484G      & 29.06                   & 0.8616                   & \textcolor{red}{0.1864}                   & 0.2277                   \\
                                                                          & {[}1, 1, 1, 1, 1, 1] & 829.3k   & 4.169G      & 29.26                   & 0.8631                   & 0.1866                   & 0.2278                   \\
                                                                          & {[}2, 2, 2]          & 621.9k   & 3.771G      & \textcolor{red}{29.54}                   & \textcolor{red}{0.8647}                   & 0.1892                   & \textcolor{red}{0.2266}                   \\
\hdashline
\multirow{3}{*}{\begin{tabular}[c]{@{}l@{}}UGSwinSR\\(ours)\end{tabular}} & 2                    & 1.203M   & 98.63M      & \textcolor{red}{27.42}  & \textcolor{red}{0.8298}  & 0.1348                   & \textcolor{red}{0.2037}  \\
                                                                          & 3                    & 4.943M   & 286.9M      & 27.13                   & 0.8236                   & 0.1298                   & 0.2082                   \\
                                                                          & 4                    & 19.85M   & 380.0M      & 26.48                   & 0.8146                   & \textcolor{red}{0.1268}  & 0.2107                   \\
\bottomrule
\end{tabular}
\end{table*}

The objective index comparison between models is also a key point to verify the advantages and disadvantages of the model. We compare the effect of the model from the aspects of parameter number, calculation amount, peak signal-to-noise ratio, structural similarity\cite{Wang2003ssim} and the learned perceptual image patch similarity (LPIPS)\cite{Zhang2018lpips}.

Table \ref{tab:different methods} shows the performance of different methods in objective indexes mentioned above. SwinIR in Table \ref{tab:different methods} has 4 RSTB, and each of which has 6 Swin Transformer Layers (STL). While the depth of MSwinSR in Table \ref{tab:different methods} is [2, 2, 2]. The reason for this comparison is that they have the same count of MSA. The former one has $4\times 6=24$ MSA, while the latter one also has $(2+2+2)\times 4=24$ MSA. Through this comparison, we can see the difference between the structure of MSwinSR and SwinIR in model complexity and reconstruction effect.

USwinSR in Table \ref{tab:different methods} represents U-net SwinSR. In other words, the discriminator is removed compared to UGSwinSR, and the loss function is $L_1$ loss.

UGSwinSR+MSwinSR in Table \ref{tab:different methods} replaces the Swin Transformer Block in Figure \ref{fig:UGSwinSR} with MSTB in Figure \ref{fig:MSwinSR}.

The following conclusions can be drawn from Table \ref{tab:different methods}:

1. The models with GAN have the highest perceptual quality (LPIPS), while SwinIR and MSwinSR have the highest objective quality (PSNR and SSIM);

2. Compared with SwinIR, MSwinSR can reduce the number of parameters and calculation amount, but there is no significant difference in other evaluation indexes. There is even a slight improvement in PSNR, which shows the advantage of the structure of MSwinSR compared with which of SwinIR;

3. UGSwinSR can effectively reduce the amount of calculation of the network, which is only about one tenth of SwinIR, and it is the lowest among several typical networks if its Swin Transformer Block is replaced by MSTB. However, the number of parameters of UGSwinSR is significantly higher, because the downsampling process will increase the number of channels of the feature map, and the number of parameters is closely related to the number of channels;

4. The combination of UGSwinSR and MSwinSR in Table \ref{tab:different methods} doesn't perform well on those metrics, which means this combination can not achieve the benefits of both the high perceptual quality of UGSwinSR and the high objective quality of MSwinSR.

Table \ref{tab:depths} shows the performance of MSwinSR and UGSwinSR under different network sizes, which shows that with the increasing size of network, the model does not necessarily perform better. In MSwinSR, it is worth noting that when depth is [1, 1, 1, 1, 1, 1] and [2, 2, 2], the number of MSA in these models are the same as 24, but the latter performs better. In UGSwinSR, the performance of the model decreases with the increase of depth, which is probably because the size of the image used in training is small, and excessive downsampling makes the network pay too much attention to local details and ignore the overall continuity of the image.

\section{Conclusion}
In this paper, we propose two main models named as MSwinSR and UGSwinSR to reduce the number of parameters and the amount of calculation. The former one uses Multi-size Swin Transformer Block (MSTB), which mainly contains four parallel multi-head self-attention (MSA) blocks to replace the residual Swin Transformer blocks (RSTB) in SwinIR\cite{liang2021swinir} . With the same number of MSA, MSwinSR performs better compared with SwinIR\cite{liang2021swinir}  in both model complexity and reconstruction effect. The latter one uses both downsampling and upsampling to decrease the calculation cost, and use GAN to reach a higher perceptual quality. But the combination of UGSwinSR and MSwinSR in this paper, which replaces the Swin Transformer Block in UGSwinSR with MSTB in MSwinSR, can not combine the advantages of both the high perceptual quality and the high objective quality. How to reduce the number of parameters of UGSwinSR and how to combine the advantages of these two models will be our further research direction.

\bibliographystyle{unsrt}  
\bibliography{references}

\end{document}